\documentclass[epjST]{svjour}
\pdfoutput=1
\usepackage{graphicx}
\usepackage{lineno}

\usepackage{cite}
\usepackage{amsmath}

\usepackage{eurosym}

\usepackage{verbatim}

\usepackage{array}
 
\begin{document}
%
\title{Analysis of renewable energy sources and electric vehicle penetration into energy systems predominantly based on lignite }
\author{
Aleksandar Dedinec\inst{1}
 \and
Borko Jovanovski\inst{1}
 \and 
Andrej Gajduk \inst{1}
 \and 
Natasa Markovska\inst{1}
 \and 
Ljupco Kocarev\inst{1,2}\fnmsep\thanks{\email{lkocarev@ucsd.edu}}
}
\institute{Macedonian Academy of Sciences and Arts, Skopje, Macedonia \and Faculty of Computer Science and Engineering, Ss. Cyril and Methodius University, Skopje, Macedonia, and Bio Circuit Institute, University of California San Diego, USA  }
\abstract{
We consider an integration of renewable energy into transport and electricity sectors through vehicle to grid (V2G) technologies for an energy system that is predominantly based on lignite. The national energy system of Macedonia is modeled using EnergyPLAN which integrates energy for electricity, transport and heat, and includes hourly fluctuations in human needs and the environment. We show that electric-vehicles can provide the necessary storage enabling a fully renewable energy profile for Macedonia that can match the country's growing demand for energy. Furthermore, a large penetration of electric vehicles leads to a dramatic reduction of 47\% of small particles and other air pollutants generated by car traffic in 2050.
%
} 
\maketitle


\section{Introduction}

The need to reduce greenhouse gases means transforming our burning-fossil-fuels society to renewable-energy-and-energy-efficiency society, which, in turn, will result in improved health, particularly through reduced air pollution. Although many countries have adopted policies to increase both energy conservation and the share of renewable energy resources, their implementations have not yet been widely adopted. Transforming its energy system traditionally based on the burning of fossil fuels to efficient, decentralized and renewable, Denmark is frontrunner in the implementation of renewable and energy efficiency policies  \cite{lund2010implementation}. In 2014, almost 40\% of the electricity consumption in Denmark was based on wind power and the plan is by 2035, 100\% of electricity and heat supply to be coved by renewable energy \cite{DenmarkRES}. Other countries/regions, however, still have to find their own strategies and approaches to deal with climate changes.     

Undoubtedly, counties/regions that have implemented or will implement renewable energy sources and/or energy efficiency technologies will have to adopt the integration of transport and electricity sectors. Transport and electric utility are two massive man-made systems for converting energy from petrochemical, nuclear, wind and solar to kinetic and electrical. Both system deliver essential energy to millions of consumers worldwide. Yet there are notable differences between these two systems: transportation is used sparingly, less then 10\% of the time~\cite{pearre2011electric}, as opposed to the average 50\% utilization of electrical power plants~\cite{eia}. Furthermore, vehicles come with large fuel tanks, while the power system lacks any significant storage. 

The increasing penetration of plug-in electric vehicles was seen as an opportunity to integrate these two complementary systems using a technology refereed to as vehicle-to-grid (V2G). This technology enables electric vehicles to take on the role of electric generators and feed power back the grid. So far V2G has been shown to: facilitate integration of intermittent energy sources~\cite{mwasilu2014electric}; provide active power regulation and spinning reserve~\cite{sortomme2012optimal}; help in peak load shaving~\cite{ota2012autonomous} and even improve grid stability~\cite{gajduk2014improving}. 

Previous research has analysed various interactions of power system and electric vehicles in various cities and countries. In urban region of Florianópolis, Brazil the pick demand energy market is analysed using the V2G concept as a grid-stabilisation strategy \cite{drude2014photovoltaics}, reduction of 785 kt $CO_2$ can be achieved with 10\% penetration of Electric Driven Vehicles (EDV) accompanied by 87 wind turbines by 2020 in Istanbul \cite{yagcitekin2013assessment}. In Japan the $CO_2$ emissions in the transport sector in 2050 can be decreased for around 81\% compared to 1990 level as a result of RES penetration and the share of both battery electric vehicles (BEV) and Fuel Cell Electric Vehicles (FCEV) which reaches around 90\% and 60\% in passenger and freight transport, respectively \cite{oshiro2015diffusion}. In Germany, a scenario with 15\% share of total vehicles to be BEV is considered in \cite{jochem2015assessing}. 100\% replacement of Latvian passenger cars with electric by 2030 could provide an important benefit for “peak shaving” in a power system with installed capacity of 2000 MW wind and 400 MW solar \cite{udrene2015role}, in Belgium BEV charging during off-peak hours instead of peak hours, can reduce $CO_2$, $SO_2$, $NO_X$ and $PM$ emission per km, significantly considering the country power generation profile \cite{rangaraju2015impacts}.


Internal combustion vehicles (ICV) are responsible for roughly 30\% of the total energy consumption and 27\% of $CO_2$ emissions in Europe~\cite{eea}. On the other hand ICVs are a major source of air pollution, e.g. in Macedonia they produce 35\% of the total NOx emissions, 27\% of particulate matter (PM $<10 \mu$G) and 17\% of other non-methane volatile organic compounds~\cite{mk}. This is caused in part by the prevalence of diesel vehicles (34\% of the vehicle fleet) and aged vehicles (73\% of all vehicles are over 10 years old) both of which contribute to high emission rates. 

The effects of transport emissions are emphasized in cities where traffic is dense and normal air dispersion is reduced. For example, in Skopje the capital of Macedonia the daily limit of 50 $\mu$G/m$^3$ PM was surpassed on roughly 200 days in 2013 whilst in Tetovo PM concentration was higher a record 308 days, reaching values of up to 800 $\mu$G/m$^3$~\cite{mk}. Long term exposures to high quantities of $PM_1$$_0$ are the cause of 6\% of total mortality, especially in respiratory deaths (15.5\%) and cardiovascular deaths (10.3\%)~\cite{kuenzli2000public,clancy2002effect}. Even medium and short term exposures have been associated with chronic bronchitis, asthma attacks, altered cardiac autonomic function, hospital admissions and restricted activity days~\cite{pope2006health}.

On the other hand, electric vehicles have zero tailpipe emissions. Shifting to electric vehicles is thus a promising way to reduce air pollution, but only if combined with renewable energy sources. In fact, according to a recent study a 10\% penetration of EV leads to 11\% reduction in direct PM-emissions, but an increase of 5-8\% in indirect emission due to higher electricity use~\cite{baumann2012contribution}. The rise in indirect emissions is caused by the coal powered plants that supply 46\% of the total energy (lignite 24\% and hard coal 22\%) in their case study. If more coal plants are replaced by wind and solar, indirect emissions will also reduce.

The complex relationships between energy, changing climate and health, on the one hand, and the lack of appropriate models for studying such relationships, on another hand, pose serious problems in developing strategies and policies of a country.   
This is the case for the country of Macedonia and this paper provides the first attempt to address these problems, namely the relationships between energy, climate change and health, for Macedonia - a country that has an energy system predominantly based on lignite.

We analyzed three separate scenarios, one baseline and two with emphasis on introducing significant percentage of PEV combined with the cost and benefits that come with them. In the second scenario we assume a moderate RES share and no V2G. Conversely, the third scenario is based on a high RES share and using electric vehicles for storage. Two separate time periods and penetration levels will be thoroughly reviewed: 2035 and 2050 with 35\% and 50\% PEV penetration, respectively.

\section{Methodology}

\begin{figure}[h!]
	
	\centering
	\includegraphics[width=0.8\textwidth]{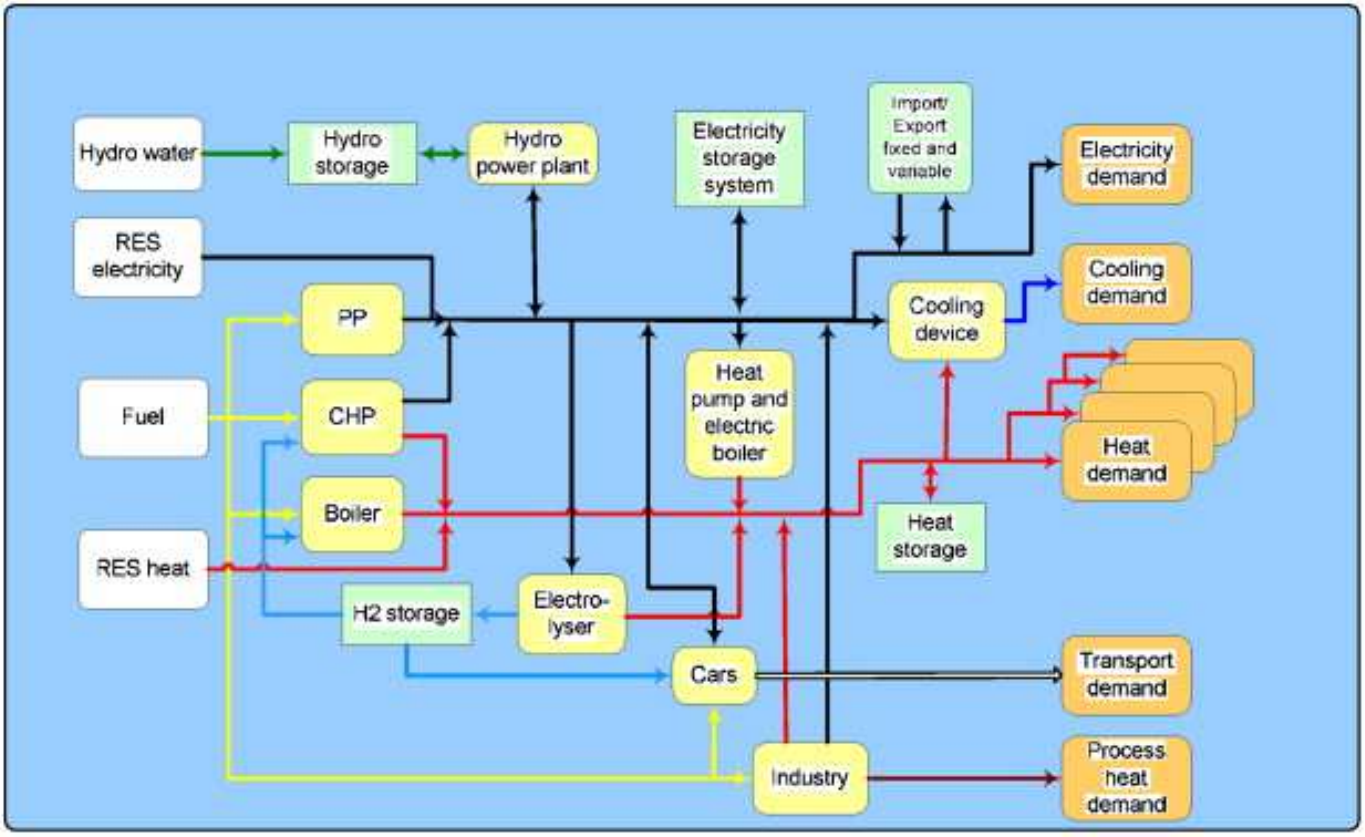}
	\caption{The EnergyPLAN energy system analysis model}
	\label{fig20}
\end{figure}

The EnergyPLAN model \cite{EnergyPlan_meth} is an analytical tool designed for analysing the energy systems on regional and national level. This model is input-output model, which uses data on capacities and efficiencies of the energy conversions of the system and availability of fuels and renewable energy inputs. Hour by hour it calculates how the electricity and heat demands of the complete system will be met under the given constraints and regulation strategies. This kind on calculations are also essential in evaluating V2G in a system that integrates a large share of renewables. Fig. \ref{fig20} illustrates the functioning of the model, showing that it covers interactions among electricity, heat and transport fuels, although it concentrates on the electrical system. This makes EnergyPLAN suitable for the combined energy system analysis needed for the investigation undertaken here \cite{lund2008integration}. 


The model uses historic data on power production rates and energy conversion systems for the power system as input and calculates how the fluctuating energy demands will be met under the specified constraints.
%
%
The model is used extensively to analyze different power system approaches: sustainability through desalination~\cite{novosel2014100}, smart energy solutions~\cite{lund2014renewable}, demand side management~\cite{brandoni2014heat}, price balancing~\cite{alagoz2013closed}.

The EnergyPLAN was used for annual analysis of the Macedonian power system because it supports the most important aspects of it: (1) large scale integration of renewables~\cite{lund2005large,connolly2010review}, (2) combined heat and power plants (CHP) \cite{lund2003modelling,lund2015large} and (3) transport integration \cite{mathiesen2009comparative}. For input we provide time series data on electricity, heating and transport energy demands and production of wind, hydro, solar, thermal and photovoltaic energy. The model also uses information on storage capacities, planned investments and regulations strategies.
Using all this data the model calculates: import costs and export revenues, critical excess costs, fuel consumption, fixed and variable operation costs, system-economic costs, $CO_2$ emissions and share of renewables.

In order to define the V2G connection to the power grid we provide the following input data: (1) electricity demand of electric vehicles in TWh/year, (2) fraction of PEV connected to the grid and available during peak hour (3) maximal power capacity of the grid connection to electric vehicles in MW (4) efficiency of the bi-directional charger/inverter between PEVs and the power grid and (5) the capacity of the battery storage in GWh.

\section{Case study: scenario definition for Macedonian energy sector }

\subsection{Sector profile}

The energy system of the Republic of Macedonia is based on fossil fuels. In the primary energy consumption 82.3\% (46.8\% solid fuels, 31.7\% oil products and 3.8\% natural gas) in 2012 were fossil fuels, 7.7\% imported electricity and the remaining 10\% renewable energy \cite{Balance}. The solid fuels share of 46.8\% makes Macedonia a country predominantly based on lignite. This energy mix makes the energy sector the main contributor of GHG with around 74\% of the total $CO_2$-eq emissions or around 80\% of the total $CO_2$ emissions \cite{Macedonia_inventory}.    

Over 90\% of the lignite is used for electricity production purposes. 70\% to 80\% (depending on the hydrology) of the electricity production in Macedonia comes from lignite power plants, and the remaining 20\% to 30\% is generated by hydro power plants \cite{MEPSO}. In the total electricity demand the import of electricity contributes with 20\% -30\% with continuous growth \cite{MEPSO}.

The total installed capacity in 2012 was 1748 MW of which 736 MW coal power plants, 198 MW heavy fuel oil power plant (several years this power plant does not work as a result of the high price of heavy fuel oil), 560 MW hydro power plants (of which 10 MW with feed-in tariffs), 280 MW gas combined heat and power (CHP) plants and 4 MW solar power plants \cite{Strategy2015}.

The transport sector in Macedonia contributed with around 25\% in the national energy balance in 2010 and the $CO_2$ emissions were around 13\% \cite{dedinec2013assessment}. The energy consumption per capita in the transport sector amounts 200 toe per 1000 inhabitants, which is more than three times lower than the EU-27 average figure (650 toe per 1000 inhabitants).
\subsection{Data and assumptions}

The input data that are used to develop the energy system of Macedonia are based on official documents and research and only a fraction of them are based on assumptions. put data used in the model are taken from energy balance \cite{Balance}, energy prices in Macedonia \cite{ERC_prices}, potential of the energy sources (lignite, hydro, biomass, etc.) inKey including import and export \cite{Strategy2015}, \cite{RES_Strategy}, installed capacities and characteristics of the power plants \cite{ELEM} and combined heat and power (CHP) \cite{TETO} and load curve for 2012 \cite{MEPSO}. The import price of electricity is at the same level as in the Energy Strategy of Maceodnia (70 Euro/MWh in 2035), \cite{Strategy2015}. 
In the Energy Plan model, in addition to the existing technologies for the production of electricity and heat a number of new technologies are introduced. For each technology the input data as: lifetime, efficiency, investment cost and fixed and variable operation and maintenance costs are defined (Tab. \ref{tab:prices}).

\begin{table}[] 
\renewcommand{\arraystretch}{1.3}
\begin{tabular}{ >{\centering\bfseries}m{0.21\hsize} |
                   >{\centering}m{0.11\hsize} |
                   >{\centering}m{0.11\hsize} |
                   >{\centering}m{0.13\hsize} | 
                   >{\centering}m{0.12\hsize} |
                   >{\centering\arraybackslash}m{0.12\hsize} }

\hline

				Power Plant Type                               & \textbf{Lifetime} & \textbf{Efficiency} & \textbf{Investment Cost (M\euro/GW)}	& \textbf{Fixed O \& M (M\euro/GW)} & \textbf{Variable O \& M (\euro/MWh) } \\ 

\hline 
\hline

Revitalization of lignite fired TPP Oslomej &	20 &	0,32  & 1211 &	25,31 &	4,6 \\
Lignite Fired & 35 &	0,40  & 		2623 &	25,31 &	4,6\\
Gas CHP* &	30 &	0,52 & 		1090 &	8,08 &	1,4\\
Gas Fired &	30 &	0,58 & 		1090 &	8,08 &	1,4\\
Hydro PP** &	50 &	& 		1300 &	1.5 &  	1\\
Wind PP &	25 &		&		1200 &	25,6 &	\\
Solar PP  &	40 &		&		800 &  	31.4 &	\\
\hline 
\end{tabular}
* Electricity efficiency for CHP plants\\
** Average price from the all hydro power projects in Macedonia \cite{Strategy2015}	
\caption{Characterization of Key Power Plant Options}	
\label{tab:prices}				
\end{table}

From the economic point of view the average rates of commercial loans in national currency (Denars) in Macedonia have decreased from 9.5\% in 2010, 8.9\% in 2011, 8.5\% in 2012, 8.2\% in 1st quarter of 2013 to 8.1 in 2nd quarter of 2013 Macedonia interest rate \cite{Interest_rate}. In this paper it is assumed that this trend of decreasing will continue so, an interest rate of 7.5\% is used.  

In the ENERGY PLAN model as an input data it is necessary to enter the electricity production of hydropower plants also at hourly level \cite{MEPSO}. Distribution curve with hourly resolution for wind was created using hourly wind speed provided by METEONORM program \cite{METEONORM}. Distribution curve with hourly resolution for solar was created using Collares-Pereira and Rabl model \cite{Collares_Pereira} and daily radiation data from \cite{NASA}.

Fuel prices, in 2013 real terms, are taken from the world energy outlook 2014, new policy scenario\cite{WEO}. In this paper new policy scenario is used because of the drop of the prices of oil products, so, this scenario now describes the current and probably the future situation more realistically.

In order to determine the total greenhouse gas emissions in the analyzed years, Energy Plan model is used in which energy consumption of all sectors (households and service, industry and transport) is inserted. The energy consumption in the other sectors is introduced in order to have more clear picture regarding the $CO_2$ emissions in Macedonia in 2035 and 2050.  Energy demand data for 2035 of these sectors is used from the draft version of the Strategy for Energy Development by 2035 \cite{Strategy2015} and for 2050 extrapolation in terms of the strategy is used and the results are adjusted to the Third National Communication (TNC) \cite{TNC}. The projections in these two documents are made using the optimization MARKAL Macedonia model \cite{MARKAL} \cite{Dedinec}. The consumption for 2050 is adjusted to the TNC because it is an official document adopted by the Government of the R. Macedonia. To calculate $CO_2$ emissions, default emission factors from the IPCC were used, while for coal country specific emission factor developed for the Second National Communication \cite{TNC}.

The type of vehicles used for the scenarios have battery capacity of 70kWh and 85kWh cause those are the batteries used by the currently most successful producer of long range electric vehicles, Tesla \cite{TESLA_motors}. In order for the vehicles to achieve the percentage of total vehicles that we present they will have to be long range and currently Tesla is setting the standard for this kind of vehicles.
According to initial data the battery degradation rate is lower than excepted and although Tesla initially assumed 30\% battery degradation after 160000 km but the data shows that the degradation is much lower, something closer to 15-20\% \cite{TESLA_capacity} and cause of the distributed approach of V2G integration the batteries will not be affected as much cause the same car will not be repeatedly used for grid stabilization.

Even in peak hours studies have shown \cite{kempton2001vehicle} that more than 80\% of the vehicles are left in the garages and in the case of EVs they are plugged to the grid leaving more than enough capacity for V2G purposes. Some EVs already have smart charger that can be set to charge the car at the cheapest rate in conjunction with signals from the electric system. This system can be easily further expanded to serve for initiating the grid stabilization procedure and using the storage for the excess power present in the grid but also pulling power from the battery if needed. It will be up to the user to decide how much of the time will let the car be available for V2G for which she will be subsidized.

\subsection{Baseline scenario}
In order to estimate the consequences of renewable energy and electric vehicles (EV) penetration in a system predominantly based on lignite a baseline scenario was developed which covers the years 2012 (base year), 2035 and 2050. In the baseline scenario, also called scenario without measures, the characteristics of the energy system in Macedonia are taken into consideration.  

In the Reference Scenario some constraints are introduced. One of the constraints is inability of construction of new large hydropower plants due to lack of investors and/or resistance of non-governmental organizations and local community. The second constraint applied to the renewable energy sources is that the capacity of power plants with feed-in tariffs (hydro, wind and solar) is limited to the capacity for which the Energy Regulatory Commission of the Republic of Macedonia has been granted at least a temporary license for preferential producer in 2014 \cite{ERC}.

Taking into account these constraints the power system in 2035 and 2050 is modeled. In 2035 it is assumed that from the existing lignite power plants of 736 MW, 626 MW will be decommissioned, and 110 MW will continue to work after revitalization and modernization. Also, the exiting CHP will be decommissioned. The lack of electricity due to decommission of the existing lignite and gas power plants in 2035 will be supplemented by 900 MW thermal power plants of which 675 MW are coal power plants, 225 are gas power plants and new gas CHPs of 240 MW. Regarding the renewable energy additional 100 MW hydropower plants, 50 MW wind power plants and 14 MW solar power plants are planned. The energy portfolio for 2035 is taken from the baseline scenario from the new Strategy for Energy Development of the Republic of Macedonia up to 2035 \cite{Strategy2015}. This portfolio is used because it is based on results obtained with optimization MARKAL Macedonia model. In 2050 the installed capacity of coal power plants is planned to be increased by 225 MW and the installed capacity of gas power plant by 75 MW. The energy portfolio for 2050 follows the same trend as 2035.

\subsection{Renewable energy sources (RES) scenario}
Renewable energy sources (RES) scenario is developed in order to highlight the advantages and disadvantages of their introduction into a system predominantly based on coal. In this scenario it is assumed that the installed capacity of coal and gas power plants remains the same as in the baseline scenario, while the installed capacity of wind power plants, solar power plants and hydropower plants increases. The installed capacity of the hydropower plants in 2035 is projected to be 1420 MW, the installed capacity of the wind power plants is 350 MW and 190 MW solar power plants \cite{Strategy2015}. The electricity consumption is the same as in the baseline scenario.

In 2050 the installed capacity of the hydropower plants will increase for 80 MW compared to 2035, so the total installed capacity would amount 1500 MW. The installed capacity of wind power plants is expected to be 650 MW, increased by 300 MW compared to 2035, and the installed capacity of solar power plants is expected to be 500 MW or 310 MW more that in 2035.

\subsection{RES + electric vehicles (RES+EV) scenario }

The role that electric vehicles may have in a system in which almost 50\% of energy is produced from renewable energy sources is recognized in the Renewable energy sources and electrical vehicles (RES + EV) scenario. The difference between this and RES scenario is in the introduction of the electrical vehicles, all other inputs are the same. In the RES + EV scenario it is assumes that 35\% and 50\% of diesel and gasoline cars will be replaced by electric cars in 2035 and 2050, respectively. 

The total number of cars is calculated using the "S" curve of motorization and the growth of GDP / capita. Thus it is obtained that the number of cars per 1000 inhabitants in 2035 is estimated at 338, increase of 190 cars per 1000 inhabitants compared to 2012, while in 2050 the number of cars per 1000 inhabitants is estimated at 462 cars. This means that there will be about 130\% more cars in 2035 and about 310\% more cars in 2050 compared to 2012. According to this the total number of cars in 2035 is estimated at 680000 in 2035, and in 2050 at 868000. The calculated number of cars in Macedonia are also similar with the number of cars in the countries of the region for 2012 that had similar GDP/capita as Macedonia is going to have in 2035 and 2050 \cite{EUROSTAT_cars}. It is assumed that up to 2035 Macedonia will increase its average number of km/year to 12000, but still lower from Slovenia in 2010 and 2014 \cite{Slovenia_km}, because of lower GDP/capita in 2035 compared to Slovenia in 2010 \cite{Strategy2015}. As a result of higher GDP/capita in 2050 compared to Slovenia in 2014 it is assumed that Macedonia is going to have 13000 km/year.

The number of EV used in this scenario are 228000 each with a battery storage capacity of 70 MWh making a total storage capacity of 15.9 GWh, and each with a 10 kW capacity of grid to battery connection. The efficiency of the battery to grid connection is 90\%. All of the vehicles considered have smart charging capabilities and it is assumed that at each moment 70\% of the total number of vehicles will be parked and connected to the grid making them extremely suitable for stabilization purposes, especially needed when a large part of the power production is from wind farms. In 2050 the total number of electric vehicles is increased to 408000 and they have an increased battery storage capacity, in comparison to those from the previous scenario, for up to 85 MWh, making the total storage capacity available to be 34.7 GWh.

\section{Results}
\subsection{Baseline scenario}
\label{baseline}
The results obtained by using the model for energy planning Energy Plan show that in order to meet the electricity demand in 2035 in the baseline scenario, the domestic production should be 9.3 TWh, while the rest 0.6 TWh should be provided by import. According to the results power plants using coal and gas will account for 69.4\% of the total demand for electricity, hydro power plants - 16.2\%, gas combined heat and power plants - 7.3\%, wind power plants 1\% and solar power plants - 0.3\%. The import of electricity is 5.9\% of the total demand. Thus the share of renewable energy in domestic production accounts for 17.5\%, an increase of 0.5\% compared to 2012. Viewed on a monthly and hourly level most of the electricity import is in January, February, March and December, reaching a maximum level of 644 MW.

\begin{figure}[h!]

  \centering
         \includegraphics[width=0.8\textwidth]{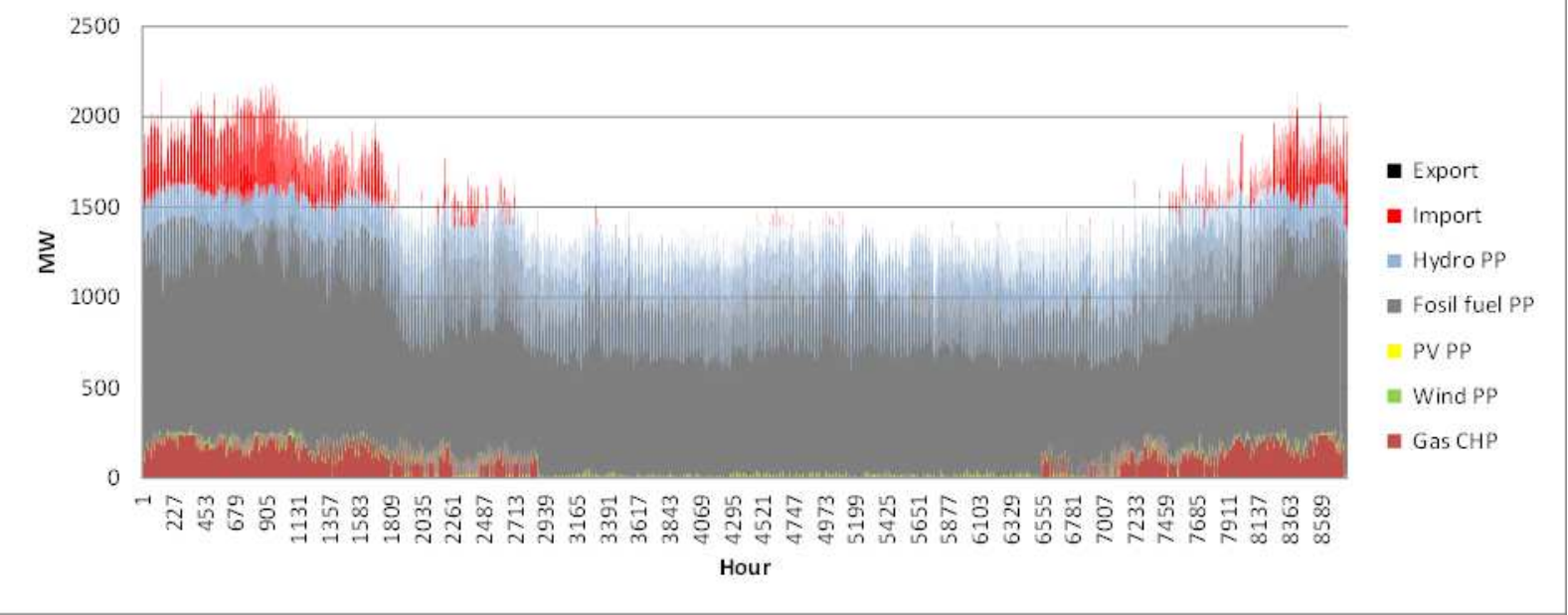}
  \caption{Production, import, export of electricity 2050}
  \label{fig1}
\end{figure}

The results show that in 2050 the electricity demand will be met by domestic production which is 11.11 TWh and 0.4 TWh from import (Fig. \ref{fig1}). Not investing in renewable energy sources in the period 2035 to 2050 in the baseline scenario contributes to the reduction of the share of renewable energy from 17.4\% in 2035 to 15\% in 2050. Coal and gas power plants in the provision of the total needs will contribute 74.9\%, an increase of 5.5\% compared to 2035, while the gas combined heat and power plants account for 6.6\%. Electricity import is reduced from about 6\% to 3.5\%. On hourly level the maximum import reaches 640 MW, and the months in which most of the import is done are the same as in 2035. During the summer period (from May up to October) almost no electricity is imported. And in the both analyzed years there is no critical exceed electricity production (CEEP), because the share of renewable energy is minimized, especially wind and solar power, so its participation did not cause the creation of surpluses in the system, viewed on hourly level.


\begin{figure}[htbp]
\begin{minipage}{0.5\linewidth}
\centering
\includegraphics[width=6cm,height=3.5cm]{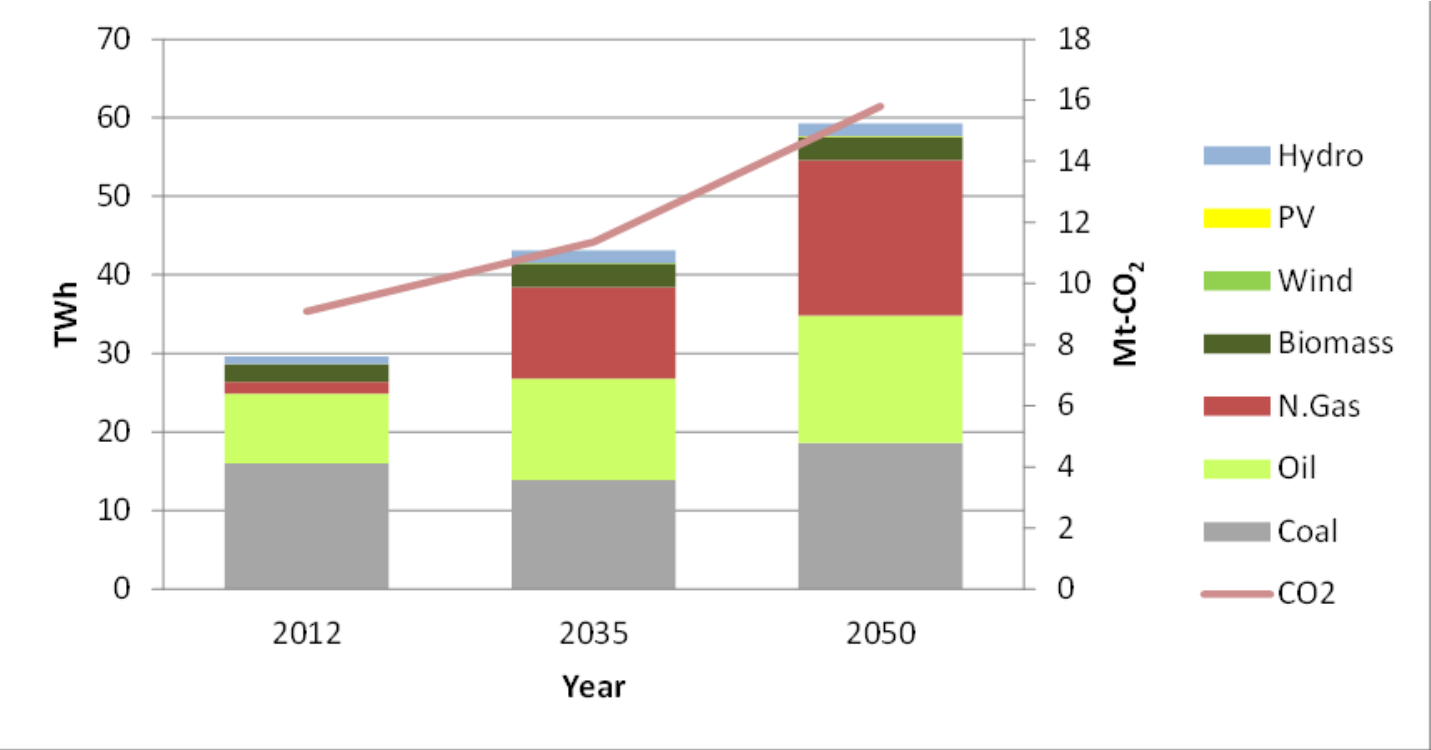}
\caption{Primary energy and $CO_2$ emission}
\label{fig2}
\end{minipage}%
\begin{minipage}{0.5\linewidth}
\centering
\includegraphics[width=7cm,height=3.7cm]{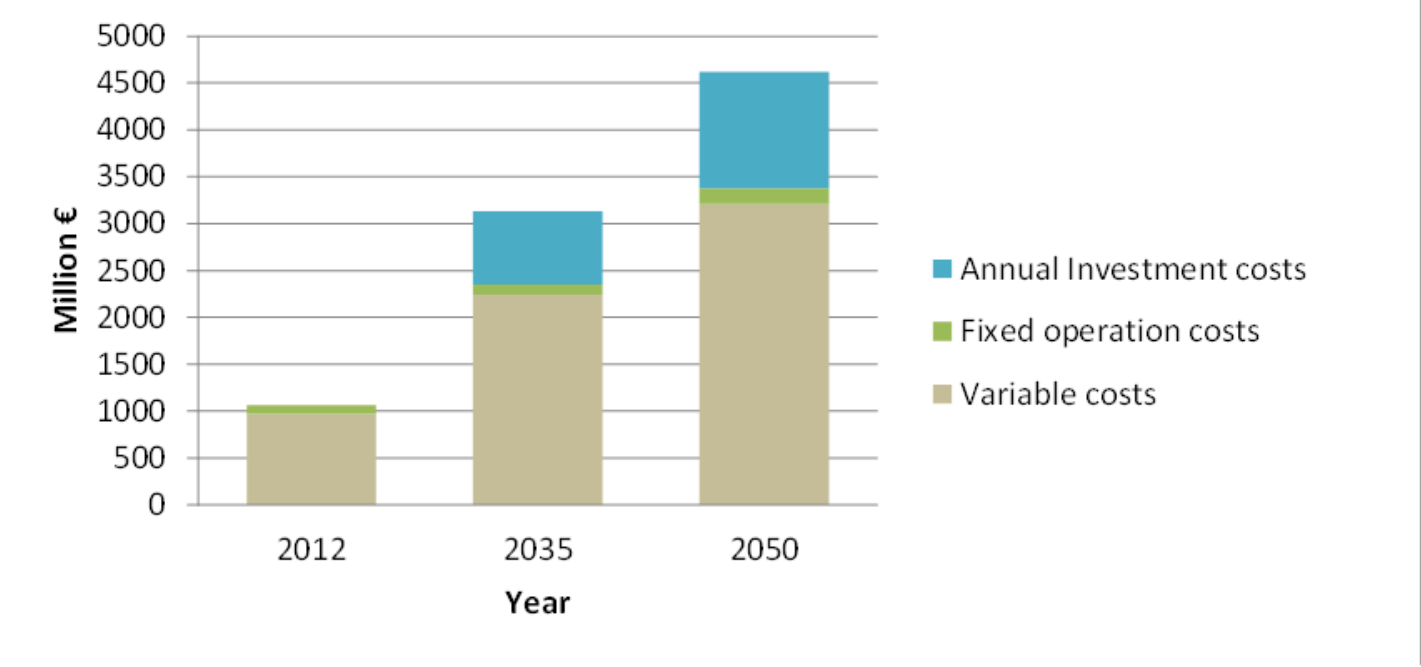}
\caption{Total annual costs}
\label{fig3}\par \medskip \vfill
\end{minipage}
\end{figure}

 As explained in the subsection data and assumption in order to determinate primary energy consumption, data for energy consumption in the other sectors in 2035 and 2050 are used. The obtained results show that the primary energy consumption will increase from 30 TWh in 2012 to 43 TWh in 2035 or about 60 TWh in 2050 (Fig. \ref{fig2}). The highest growth was recorded for natural gas and in 2035 it will contribute about 27\% and in 2050 it increases its share in primary energy consumption to 33\%. This increase is due to the gasification of Macedonia, and the construction of new gas thermal power plants. Besides gas share of 32\% in the primary energy consumption, in 2035 and 2050 coal share decreases compared to 2012 when its share was 54\%. The reduction of coal consumption is due to the decommission of the coal fired thermal power plants that have an efficiency of about 30\%, which is accompanied by opening of new ones with a much higher efficiency of about 40\%.

Using this primary energy consumption, the results show that greenhouse gas emissions in 2035 will amount to 11368 $kt$ $CO_2$, which represents an increase of 25\% compared to 2012 (9100 $kt$ $CO_2$), while in 2050, greenhouse gas emissions are projected to increase by additional 39\% compared to 2035 and will amount to 15800 $kt$ $CO_2$ (Fig. \ref{fig2}).

Using the Energy Plan model the total cost of the energy system in 2012 is estimated to be \euro 1060 million, while the overall cost of the system in 2035 is obtained to be about 3 times greater (around \euro 3100 million) compared to 2012, and in 2050 it is 1.5 times greater than in 2035 (around \euro 4600 million), as shown in Fig. \ref{fig3}. The overall cost of the energy system includes the annual cost of investments (power plants and vehicles), fixed and variable operation and maintenance costs and cost of the fuels in all sectors. Because there is investment in new coal, gas, hydro, wind and solar power plants in 2035, investment costs amount to approximately \euro 785 million or 25\% of the total costs, while in 2050 they are increased to \euro 1242 million and amount to 27\% of the total costs.

\subsection{RES AND RES + EV scenarios}
In the RES scenario the increase of the installed capacity of renewable energy sources will increase their production, so in 2035 they participate with 44.6\% in the total production which is an increase of 26\% compared to the baseline scenario. At the same time, the share of coal and gas thermal power plants is reduced by about 20\% compared to the baseline scenario and it is predicted to account for 48\%.

\begin{figure}[h!]

  \centering
         \includegraphics[width=0.9\textwidth]{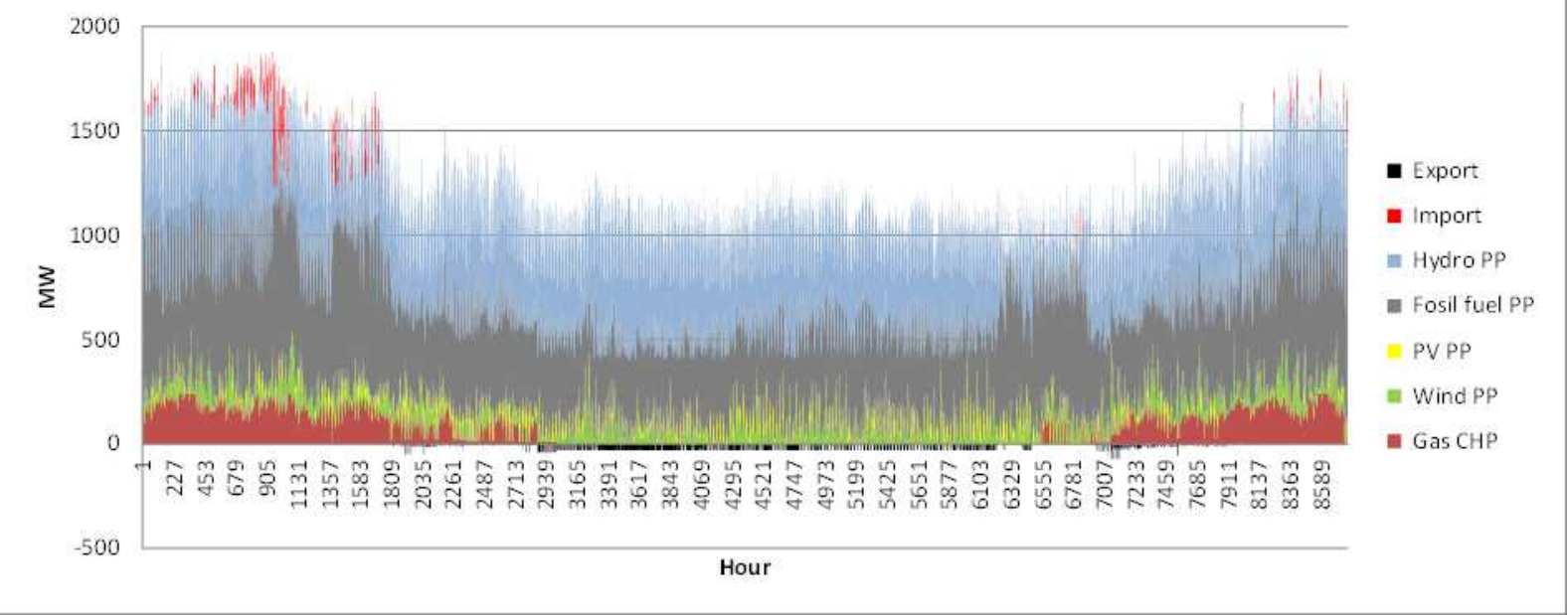}
  \caption{Production, import and export of electricity 2035 – RES}
  \label{fig4}
\end{figure}

The greater use of renewable energy sources reduces the import that amounts to 0.4\%, which is a decrease of 5.6\% compared to the baseline scenario. However, as a result of the introduction of RES in the system there is emergence of critical exceed electricity production (CEEP), or at particular times the production of electricity is greater than the demand, leading to system instability. CEEP in 2035 is projected at around 86 GWh but it is notable that in some hours this CEEP is up to 75 MW (Fig. \ref{fig4}). CEEP in general occurs throughout the year, but are most critical in the summer period, i.e. from May to September. As it is presented on Fig. \ref{fig4}, during this period the production from wind and solar power plants is increased, while the need for electricity is decreased.

\begin{figure}[h!]

  \centering
         \includegraphics[width=0.9\textwidth]{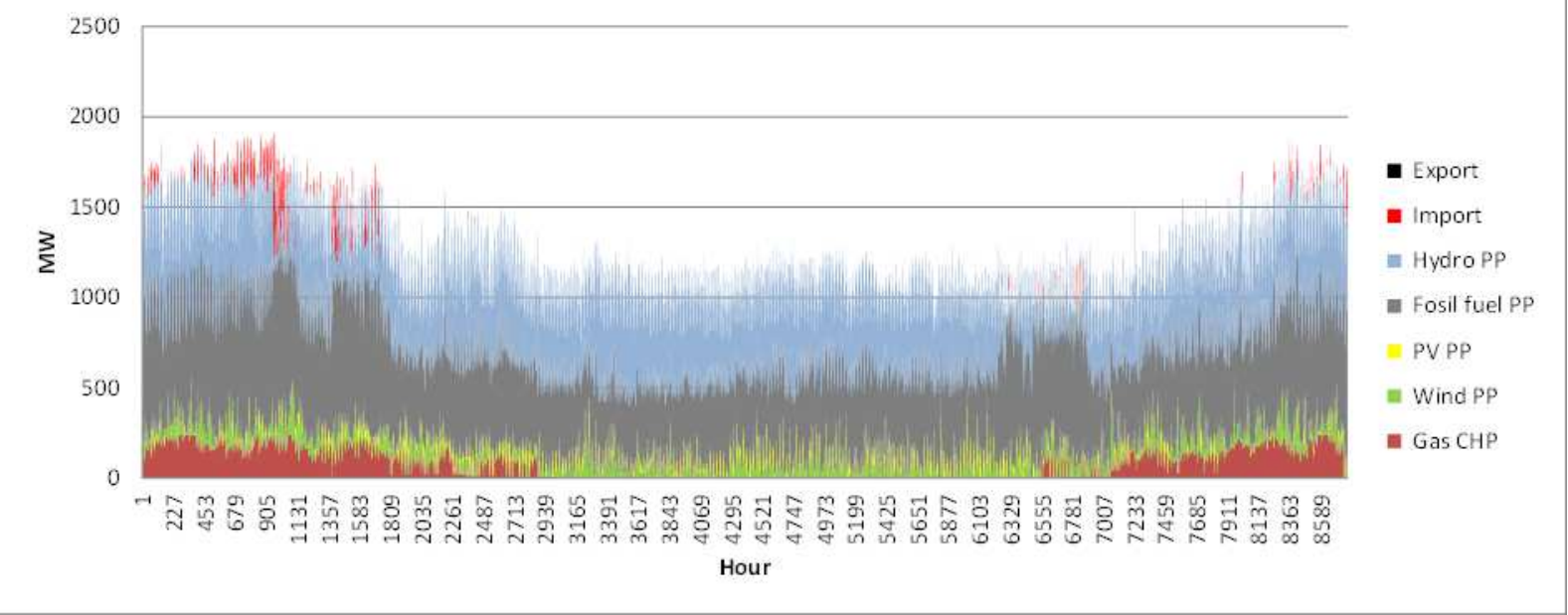}
  \caption{Production, import and export of electricity 2035 – RES+EV scenario}
  \label{fig5}
\end{figure}

In order to eliminate CEEP and ensure a proper working of the system in which nearly half of the electricity production is from coal and gas thermal power plants and half from renewable energy sources, in this paper the introduction of electric vehicles is considered. This paper assumes that in 2035 35\% of the cars will be replaced by electric vehicles. The introduction of electric vehicles contributed to reversal of CEEP, but as a result of the increased consumption of electricity, the production from coal and gas thermal power plants has increased by 1\%, and also the import of electricity has increased by 0.2\%. Besides the cancelation of CEEP, the introduction of electric vehicles improves system operation. This is shown on Fig. \ref{fig5} where in the period from May to September (when the maximum CEEP was 75 MW in the RES scenario) in the RES + EV scenario the batteries of the electric vehicles are used to over 500 MW in certain hours.

\begin{figure}[h!]

  \centering
         \includegraphics[width=0.9\textwidth]{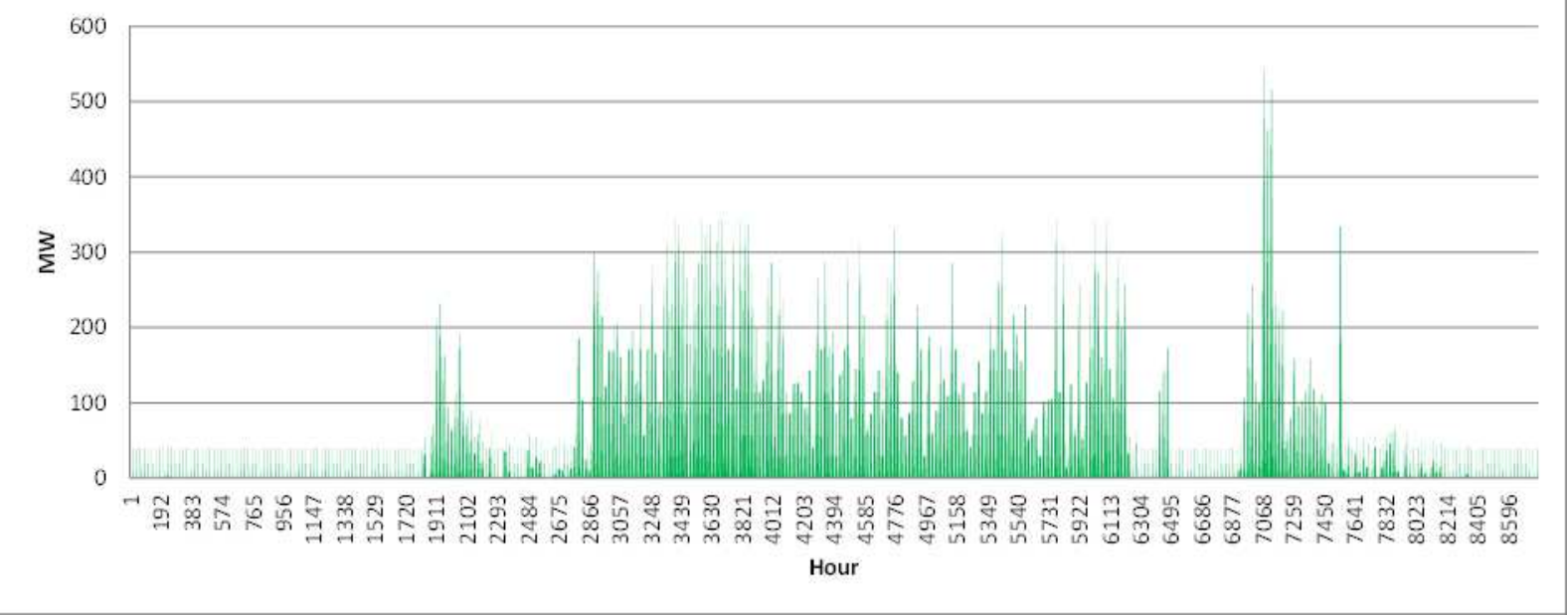}
  \caption{Storage of electricity – vehicle to grid 2035 RES+EV scenario}
  \label{fig6}
\end{figure}

The production from renewable energy sources in 2050 increases to 47.5\%, which is an increase of about 3\% compared to 2035, while the production from coal and gas thermal power plants is reduced to 46\%, which is a decrease of 2\% compared to 2035. Electricity import is reduced from 3.5\% in the baseline scenario to 0.3\% in the RES scenario.

As in 2035, the greater use of renewable energy sources in 2050 leads to the emergence of CEEP, or at particular hours greater production than electricity demand, leading to system instability. CEEP in 2050 is predicted to be around 270 GWh, reaching a maximum of 470 MW (Fig. \ref{fig6}). Overall CEEP occurs throughout the year but, again, the most critical is the period of May to September. 


50\% electric vehicle in 2050 will contribute to almost complete cancellation of CEEP. There are only five hours in the year that still have CEEP. This excess of electricity production can be exported or it can be further canceled by increasing the number of electric vehicles.
As in 2035, the introduction of electric vehicles will contribute to better performance of the system and use of storage for more than 4000 MW.



\begin{figure}[h!]

  \centering
         \includegraphics[width=0.5\textwidth]{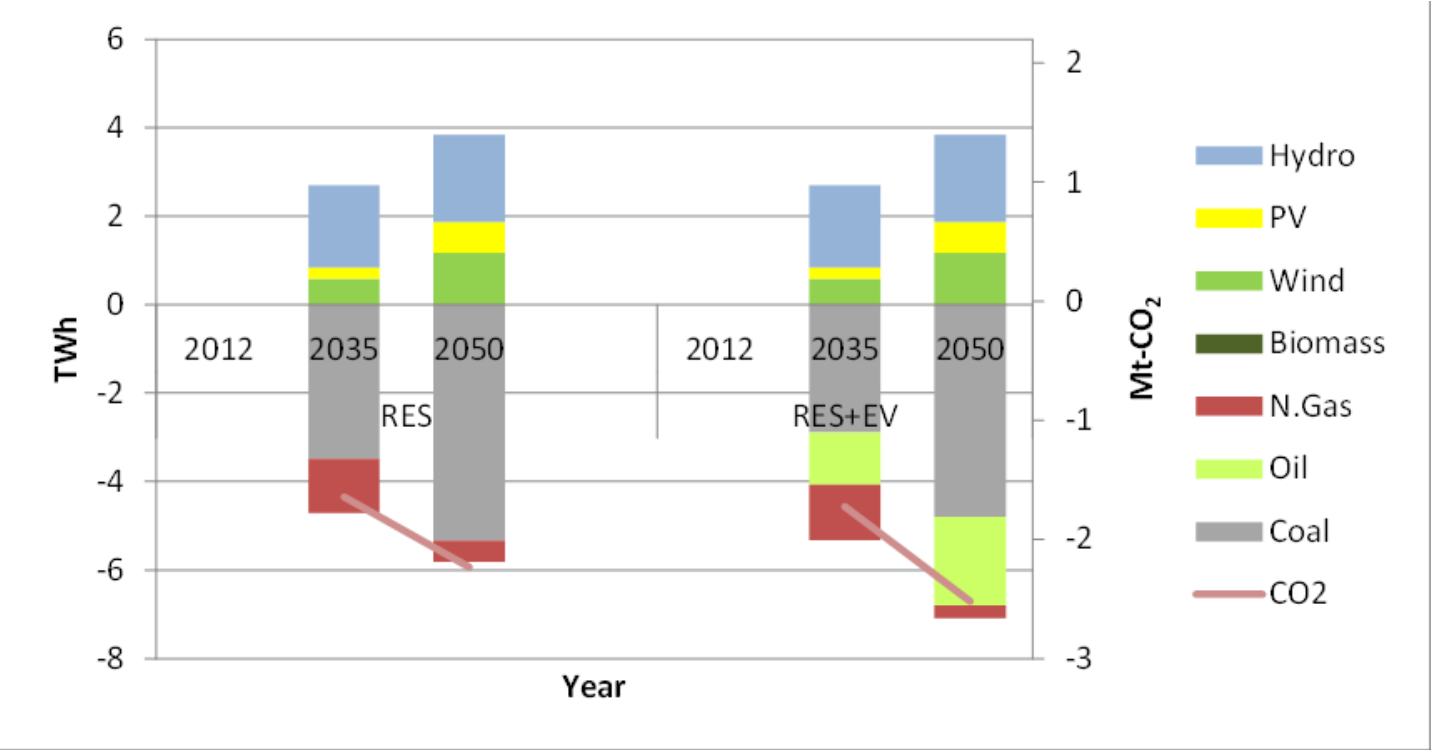}
  \caption{Primary energy and $CO_2$ emissions RES and RES+EV scenario compared to Baseline}
  \label{fig7}
\end{figure}

The primary energy in the RES scenario is reduced by about 2 TWh in both 2035 and 2050 compared to the baseline scenario (Fig. \ref{fig7}). The largest decrease occurs in coal - by 3.5 TWh and 5.3 TWh in 2035 and 2050, respectively. There is increase compared to the baseline scenario of the production of hydro power plants by about 2 TWh in 2035 and 2050, wind power for about 0.58 TWh in 2035 and 1.17 TWh in 2050, solar power plants for about 0.25 TWh, and 0.7 TWh in 2035 and 2050, respectively. In the RES + EV scenario the primary energy is decreased by 2.6 TWh and 3.3 TWh, which compared to the RES scenario, has further oil products reduction as a result of the replacement of vehicles that use fossil fuels to electric vehicles. In this scenario, compared to RES scenario there is a slight increase of coal as a result of higher electricity consumption.

Emissions of greenhouse gases in the RES scenario are reduced by 1.6 $Mt$ $CO_2$ in 2035 and 2.2 $Mt$ $CO_2$ in 2050 (Fig. \ref{fig7}). In the RES + EV scenario, in 2035, there is emissions reduction of 1.7 $Mt$ $CO_2$ in 2035 and 2.5 $Mt$ $CO_2$ in 2050 compared to the baseline scenario. Despite the higher electricity demand in the RES+EV scenario the $CO_2$ emissions are lower compared to the RES scenarios, because of the efficiency of the EV, which reduce the consumption of the oil products and on the other hand electricity which is exported in the RES scenarios, in the RES+EV scenarios is used for EV.


Another benefit of electric vehicles is the reduction of the PM. In order to calculate the reduction of PM, the data for the number of electric vehicles, the average number of km per vehicle and the emission factors are used. In the RES + EV scenario it was specified that 35\% of the cars will be electrical vehicles in 2035, which is equal to 228000 electric vehicles, while in 2050 this percent is 50\% or there will be 408000 electric vehicles. In this paper, the PM emission factor from the highest EU standard at the moment \cite{PM_directive} for fossil fuel cars is used and is equal to 0.005 g / km. Using all these data it is obtained that electric vehicles will contribute to PM reduction of 13.7 t (which is 34\% of the total car PM emissions) and 26.5 t (which is 47\% of the total car PM emissions) in 2035 and 2050, respectively. 

In order to quantify the health effects of this, a model is required to convert the gross reduction in $PM_1$$_0$ pollution to actual daily concentrations of $PM_1$$_0$ in major cities such as Tetovo and Skopje. Unfortunately, such a model requires meteorological data which to the best of our knowledge is not recorded for these cities. As a result exact quantification is difficult and is thus left for future study. Nevertheless, curbing transport based air pollution, which is known to be one of the main sources of $PM_1$$_0$ and is mainly concentrated in large cities, by almost half will reduce mortality rates and prevent pulmonary diseases.

\begin{figure}[h]

  \centering
         \includegraphics[width=0.5\textwidth]{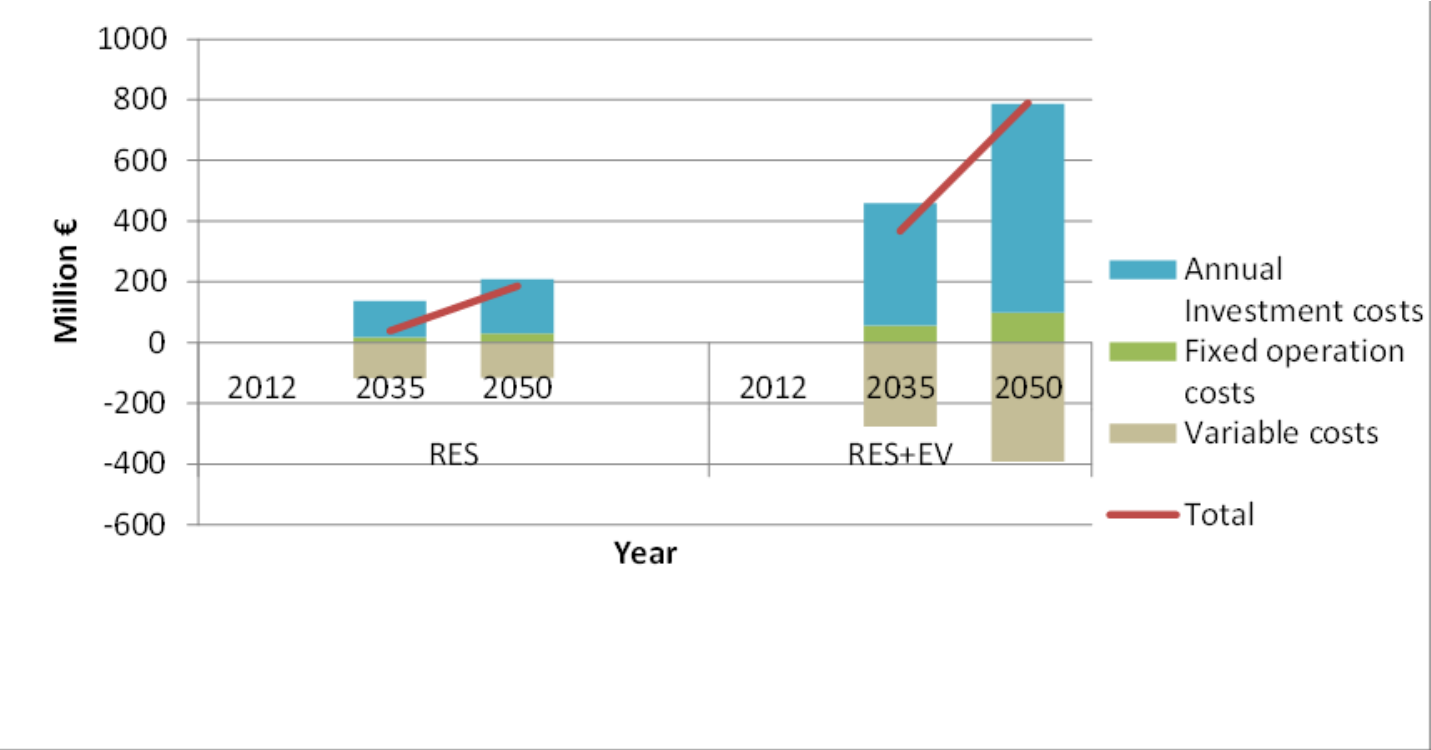}
  \caption{Total annual costs RES and RES+EV scenario}
  \label{fig8}
\end{figure}

In terms of overall costs, in both scenarios there is an increase of the total costs, by \euro 38 million in 2035 and by \euro 186 million in 2050, in the RES scenario (Fig. \ref{fig8}). The introduction of electric vehicles further increases the cost of the system and contributes to total increase of \euro 368 million in 2035 and \euro 790 million in 2050 compared to the cost of the system in the baseline scenario.

\section{Conclusion}
Reduction of GHG and PM emissions is one of Macedonia's main concerns, a country currently struggling with extreme pollution in its largest cities. This task is getting more and more difficult because through the years Macedonia has been facing the problem of electricity production mainly based on lignite power plants and old vehicles fleet.    

In this paper we have considered two different scenarios for transforming the energy system predominantly based on lignite by introducing large percentage of RES in the energy mix and have compared them to a baseline scenario that keeps the current mix unchanged. Additionally, in one of the scenarios we considered penetration of PEV, that through the concept V2G, can be utilize for grid stabilization. 

The results have shown that large percentage of PEV from the total number of passenger cars contributes to elimination of CEEP allowing greater penetration of renewable energy sources, which in turn reduces the import dependency. This represents a significant contribution towards one of the key national strategic priorities in the energy sector. From the economic point of view the introduction of RES will increase the cost of the system of 2\% in 2050, and with penetration of PEV the cost will be increased by additional 7\% compared to the baseline scenario. However, the increase in costs can be compensated to some extent by the newly created jobs associated with the higher penetration of RES. Additional study is needed to monetize this positive impact to the economy. 

Furthermore, the RES scenarios have presented numerous side benefits one of which is the considerable reduction of the $CO_2$ emissions by 1.6 $Mt$ $CO_2$ in 2035 and 2.2 $Mt$ $CO_2$ in 2050 for the RES scenario and 1.7 $Mt$ $CO_2$ in 2035 and 2.5 $Mt$ $CO_2$ in 2050 for the RES +EV scenario compared to the baseline scenario. 

Finally, we have shown that RES + EV scenario leads to a reduction in direct $PM_1$ $_0$ generated by transport by 34\% and 47\% in 2035 and 2050, respectively.  Because transport is responsible for 17\% of total PM released in the atmosphere, this reduction will undoubtedly help people with chronic illness and possibly even reduce pollution caused mortality rates.

\end{document}